# VLASOV EQUATION CORRECT APPLICATION BASED ON THE HIGHER ORDERS KINEMATIC VALUES FOR THE DISSIPATIVE SYSTEMS DESCRIPTION

E.E. Perepelkin[1], B.I. Sadovnikov[1], N.G. Inozemtseva[2], D.A. Suchkov[1]

[1]*Faculty of Physics, Moscow State University, Moscow, 119992 Russia*
[2]*Dubna State University, Universitetskaya st.19, Dubna, Moscow region, 141980 Russia*

**Abstract**

This paper revises traditional phenomenological approaches to the application of Vlasov equation to describe the dissipative systems. The original equation by A.A. Vlasov obtained differs from the classical Vlasov equation used in the scientific literature. The classical Vlasov equation cannot be used to describe the dissipative systems. The original Vlasov equation contains a non-zero right-hand side, derived from the first principles. It is shown that the original Vlasov equation describes the dissipative systems by the non-phenomenological way. The numerical modeling of the dissipative systems using the motion equations solution is performed. The numerical results show the good agreement with the exact solutions of the original Vlasov equation for the dissipative systems.

In this way, a wide range of the statistical physics problems, the plasma physics, the astrophysics, the high-energy physics, the controlled fusion using the original Vlasov equation can be revised.

**Introduction**

As of today, the Vlasov equation [1, 2] to describe a wide class of the interactive particle systems is used. Such systems examples are found in the statistical physics [3-6], the plasma physics, the thermonuclear fusion problems [7-11], the accelerator physics [12-14], the astrophysics [15-20] and the condensed matter physics [21-23]. There are the large numbers of the articles which the numerical solution of the Vlasov, Vlasov-Poisson and Vlasov-Maxwell equations are devoted. Here are a few of them [24-29].

Usually the phenomenologically modified Vlasov equations for the dissipative systems are used [30, 31]. As a rule, the right-hand equation side by the semi-phenomenological way is changed. For example, the paper [30] considers the Enskog-Vlasov equation for studying the phase transitions. The right-hand side of the Enskog-Vlasov equation contains the collision integral phenomenologically introduced. It is possible to explain the number of phenomena by using the Enskog-Vlasov equation. This phenomena are not possible explain with using the classical Vlasov equation. In [31], for the particular case of the Enskog-Vlasov equation, the H-theorem is proved. Also in [31], a comparison of the results of numerical simulation of fluid flow with experimental results is considered. The disadvantage of the Enskog-Vlasov equation is its semi-phenomenological nature and the need to specifically select the collision integral on the right-hand side of the equation.

Changes to the Vlasov equation by introducing the collision integral into the right-hand side contradict the A.A. Vlasov idea on the description of a collisionless particle system [1, 2]. The problem of inaccurate description of dissipative systems using the classical Vlasov equation consists in the incorrect form of using this equation. In [32], we showed that there is a difference between the classical Vlasov equation, which is used in the description of dissipative systems, and the original equation, which was written by A.A. Vlasov.

From the first principles, A.A. Vlasov wrote not one equation, but an infinite chain of self-linking equations for the distribution functions $f_1(\vec{r},t), f_2(\vec{r},\vec{v},t), f_3(\vec{r},\vec{v},\dot{\vec{v}},t),\ldots$ of random variables: coordinates, velocities, accelerations and higher orders accelerations. The classical Vlasov equation is understood as the second equation from the chain of Vlasov equations of the form:

$$\frac{\partial f_2}{\partial t} + (\vec{v}, \nabla_r f_2) + (\langle \dot{\vec{v}} \rangle, \nabla_v f_2) = 0, \qquad (i.1)$$

where $f_2 = f_2(\vec{r},\vec{v},t)$ is a distribution function of the random variables: a coordinate and velocity. The variable $\langle \dot{\vec{v}} \rangle = \langle \dot{\vec{v}} \rangle(\vec{r},\vec{v},t)$ corresponds to the average acceleration, that is [1,2]

$$f_2(\vec{r},\vec{v},t)\langle \dot{\vec{v}} \rangle(\vec{r},\vec{v},t) \stackrel{\text{det}}{=} \int_{(\infty)} \dot{\vec{v}}\, f_3(\vec{r},\vec{v},\dot{\vec{v}},t)\, d^3\dot{v}, \qquad (i.2)$$

where $f_3(\vec{r},\vec{v},\dot{\vec{v}},t)$ is a distribution function of the random variables: a coordinate, velocity and acceleration. For average acceleration (i.2), A.A. Vlasov used the approximation $\langle \dot{\vec{v}} \rangle = \vec{F}/m$, where $\vec{F}$ is a force actin on a particle with a mass $m$.

The original second equation in the Vlasov chain is of the form [1,2]:

$$\frac{\partial f_2}{\partial t} + \operatorname{div}_r [f_2 \vec{v}] + \operatorname{div}_v [f_2 \langle \dot{\vec{v}} \rangle] = 0. \qquad (i.3)$$

In the particular case, at $\langle \dot{\vec{v}} \rangle = \langle \dot{\vec{v}} \rangle(\vec{r},t)$ (that is $\operatorname{div}_v \langle \dot{\vec{v}} \rangle = 0$) the equation (i.3) goes into the classical Vlasov equation (i.1). In [32] we considered the «new» modified Vlasov equation

$$\frac{\partial S_2}{\partial t} + (\vec{v}, \nabla_r S_2) + (\langle \dot{\vec{v}} \rangle, \nabla_v S_2) = -Q_2, \qquad (i.4)$$

$$S_2 \stackrel{\text{det}}{=} \ln f_2, \quad Q_2 \stackrel{\text{det}}{=} \operatorname{div}_v \langle \dot{\vec{v}} \rangle.$$

The equation (i.4) is completely equivalent to the original equation (i.3) and in the general case of considering dissipative systems has a nonzero right-hand side $Q_2 \neq 0$. Unlike the Enskog-Vlasov equation, the right-hand side $Q_2$ is present in the equation (i.4) from first principles without the use of phenomenology.

A natural question arises: «How well does the equation (i.4) describes dissipative systems»? The answer to this question is the subject of the given paper.

In this paper, we consider several types of the dissipative systems model, for which we obtain the exact solutions of the equation (i.3), (i.4). Next, we perform a simulation of the dissipative systems by numerically solving the motion equations. According to the obtained data, we construct the distribution functions $f_1(\vec{r},t), f_2(\vec{r},\vec{v},t), f_3(\vec{r},\vec{v},\dot{\vec{v}},t)$, with the help of which we determine the average value of $\langle \dot{\vec{v}} \rangle$ and verify the correctness of the classical Vlasov equation (i.1) and the modified (original) Vlasov equation (i.3), (i.4).

The paper has the following structure. In §1, three types of the dissipative systems are considered. For these systems, the exact solutions of the equation (i.1), (i.4) are constructed by the characteristics, and their properties are also considered. In §2, a numerical simulation of a system of $2^{23}$ particles is performed. Numerical integration of the motion equations is performed on the massively parallel computing architecture of GPU graphics processors. Using a number of statistical criteria, in §2, a test of hypotheses is made regarding the agreement of the simulation data with the exact solutions of the equations (i.4) and (i.1). Section 3 contains a discussion of the data obtained from a numerical experiment, comparing them with exact solutions of the equations (i.1) and (i.4). In conclusion, the main results and conclusions of the paper are presented.

**§1 Exact solutions**

The infinite Vlasov equation chain can be written in a compact form [1, 2, 32]:

$$\Pi_n S_n = -Q_n, \quad n \in \mathbb{N}, \qquad (1.1)$$

where

$$\Pi_n \stackrel{\text{det}}{=} \frac{d_n}{dt} \stackrel{\text{det}}{=} \frac{\partial}{\partial t} + (\dot{\vec{r}}, \nabla_r) + \ldots + \left(\stackrel{(n-1)}{\vec{r}}, \nabla_{\stackrel{(n-2)}{r}}\right) + \left(\left\langle \stackrel{(n)}{\vec{r}} \right\rangle, \nabla_{\stackrel{(n-1)}{r}}\right), \qquad (1.2)$$

$$S_n(\vec{\xi}_n, t) \stackrel{\text{det}}{=} \ln f_n(\vec{\xi}_n, t), \quad Q_n(\vec{\xi}_n, t) \stackrel{\text{det}}{=} \text{div}_{\stackrel{(n-1)}{r}} \left\langle \stackrel{(n-1)}{\vec{v}} \right\rangle (\vec{\xi}_n, t), \, \vec{\xi}_n = \left\{ \vec{r}, \vec{v}, \ldots, \stackrel{(n-1)}{\vec{r}} \right\}^T.$$

The distribution functions $f_1(\vec{r},t), f_2(\vec{r},\vec{v},t), f_3(\vec{r},\vec{v},\dot{\vec{v}},t), \ldots$ satisfy the conditions [1, 2]:

$$f_0(t) \stackrel{\text{det}}{=} N(t) = \int_{(\infty)} f_1(\vec{r},t) d^3r = \int_{(\infty)}\int_{(\infty)} f_2(\vec{r},\vec{v},t) d^3r d^3v = \ldots =$$
$$= \ldots = \int_{(\infty)}\int_{(\infty)}\int_{(\infty)} \ldots f_\infty(\vec{r},\vec{v},\dot{\vec{v}},\ldots,t) d^3r d^3v d^3\dot{v}\ldots, \qquad (1.3)$$

Where the function $N(t)$ corresponds to the number of particles. The average values of the kinematic variables are determined by the ratios

$$N_1(\vec{r},t)\langle\langle \vec{v} \rangle\rangle(t) = \int_{(\infty)} f_1(\vec{r},t)\langle \vec{v} \rangle(\vec{r},t) d^3r = \int_{(\infty)}\int_{(\infty)} f_2(\vec{r},\vec{v},t) \vec{v} d^3r d^3v, \qquad (1.4)$$

$$N_1(t)\langle\langle\langle \dot{\vec{v}} \rangle\rangle\rangle(t) = \int_{(\infty)} f_1(\vec{r},t)\langle\langle \dot{\vec{v}} \rangle\rangle(\vec{r},t) d^3r = \int_{(\infty)}\int_{(\infty)} f_2(\vec{r},\vec{v},t)\langle \dot{\vec{v}} \rangle(\vec{r},\vec{v},t) d^3r d^3v =$$
$$= \int_{(\infty)}\int_{(\infty)}\int_{(\infty)} f_3(\vec{r},\vec{v},\dot{\vec{v}},t) \dot{\vec{v}} d^3r d^3v d^3\dot{v},$$

…

For the chain (1.1), the representation via the generalized Boltzmann $H_n$-function is correct [32]

$$\frac{d}{dt}\left[N(t)H_n(t)\right] = -N(t)\langle...\langle Q_n\rangle...\rangle(t), \qquad (1.5)$$

where

$$H_1(t) \stackrel{det}{=} \frac{1}{N} \int_{(\infty)} f_1(\vec{r},t)\ln f_1(\vec{r},t)d^3r = \frac{1}{N}\int_{(\infty)} f_1(\vec{r},t)S_1(\vec{r},t)d^3r = \langle S_1\rangle(t), \qquad (1.6)$$

$$H_2(t) \stackrel{det}{=} \frac{1}{N}\int_{(\infty)}\int_{(\infty)} f_2(\vec{r},\vec{v},t)\ln f_2(\vec{r},\vec{v},t)d^3rd^3v = \langle\langle S_2\rangle\rangle(t),$$

...

$$H_n(t) \stackrel{det}{=} \langle...\langle S_n\rangle...\rangle(t),$$

...

In the case $n=2$, the generalized Boltzmann $H_n$-function corresponds to the classical Boltzmann H-function (1.6). The functions $Q_n$ have the physical significance as sources of dissipation determined by the kinematic values of the $n$-order $\left\langle\stackrel{(n-1)}{\vec{v}}\right\rangle$. From the equation (1.5) it follows that if the dissipation sources $\langle...\langle Q_n\rangle...\rangle$ are absent, then the generalized Boltzmann $H_n$-function is constant, which can correspond to a equilibrium system. According to the equation (1.5), the sign of the sources $\langle...\langle Q_n\rangle...\rangle$ corresponds to of the increase and decrease of the Boltzmann $H_n$-function.

**Remark 1**

There are cases when the function $Q_n \neq 0$ (the dissipative system) and the average value $\langle...\langle Q_n\rangle...\rangle = 0$. In this case, despite the dissipative nature of the system, its Boltzmann $H_n$-functions is constant. An example of such a system will be discussed below.

At $n=2$, the equation (1.1) transform into the equation (i.4), and the corresponding equation (1.5) at a constant number of particles $N = const$ is of the form:

$$\frac{d}{dt}H_2(t) = -\langle\langle Q_2\rangle\rangle(t), \qquad (1.7)$$

$$\langle\langle Q_2\rangle\rangle(t) = \frac{1}{N}\int_{(\infty)}\int_{(\infty)} f_2(\vec{r},\vec{v},t)Q_2(\vec{r},\vec{v},t)d^3rd^3v.$$

In the case of classical Vlasov equation (i.1), there are no sources $Q_2$ ($\langle\langle Q_2\rangle\rangle = 0$), and $H_2$ is constant, that is

$$\frac{d}{dt}H_2(t) = 0. \qquad (1.8)$$

We show that when considering dissipative systems, it is necessary to use the modified (original) Vlasov equation (i.4) and the equation (1.7) and not the classical Vlasov equation (i.1) and the equation (1.8). We will check this statement numerically and analytically. We consider a model set of dissipative systems, for which we will find exact and numerical solutions. We will seek solutions by directly integrating the motion equations. We substitute the obtained solutions into the equations (1.7) - (1.8) and verify their implementation.

For a numerical estimate of the average acceleration $\langle \dot{v} \rangle$, knowledge of the function $f_3(\vec{r}, \vec{v}, \dot{\vec{v}}, t)$ (i.2) is necessary. The domain of the function $f_3(\vec{r}, \vec{v}, \dot{\vec{v}}, t)$ has a dimension $9D+1D$. The numerical solution of the initial-boundary value problem in space with a dimension $10D$ requires significant computational costs. Therefore, we consider the function $f_3(x, v, \dot{v}, t)$ for which the domain has a dimension $3D+1D$.

As a model system, we consider an oscillator with different types of attenuation. The motion equations are:

$$m\dot{v} = -kx, \tag{1.9}$$

$$m\dot{v} = -kx - bv, \tag{1.10}$$

$$m\dot{v} = -kx - \lambda v^2, \tag{1.11}$$

$$m\dot{v} = -kx - A\cos(\gamma v), \tag{1.12}$$

where $m, k, b, \lambda, A, \gamma$ are some constant values. The equation (1.9) corresponds to the usual harmonic oscillator, in which there are no dissipations.

The equation (i.4) for the function $f_2(x, v, t)$ is of the form

$$\frac{\partial S_2}{\partial t} + v \frac{\partial S_2}{\partial x} + \langle \dot{v} \rangle \frac{\partial S_2}{\partial v} = -Q_2 = -\frac{\partial \langle \dot{v} \rangle}{\partial v}, \quad S_2 = \ln f_2. \tag{1.13}$$

According to (1.9)-(1.12) and (i.2), the function $\langle \dot{v} \rangle(x, v, t)$ in the equation (1.13) is as follows:

1. $m\langle \dot{v} \rangle = -kx, \quad Q_2 = 0,$

2. $m\langle \dot{v} \rangle = -kx - bv, \quad Q_2 = -\dfrac{b}{m},$ (1.16)

3. $m\langle \dot{v} \rangle = -kx - \lambda v^2, \quad Q_2 = -2\dfrac{\lambda}{m}v,$

4. $m\langle \dot{v} \rangle = -kx - A\cos(\gamma v), \quad Q_2 = \dfrac{A\gamma}{m}\sin(\gamma v).$

Let us construct the solutions of the equation (1.13) for the cases (1.16) by the characteristics method. To the equation (1.13), the following characteristic equations correspond:

$$\frac{dt}{1} = \frac{dx}{v} = \frac{dv}{\langle \dot{v} \rangle} = -\frac{dS_2}{Q_2}. \tag{1.17}$$

If $C_1, C_2, C_3$ are first integrals of the equation (1.13), then the general solution of the equation can be written in the form $\Phi(C_1, C_2, C_3) = 0$, where $\Phi$ is some function [42]. Let us find the solutions (1.17) for the 1-3 cases from (1.16) (see Appendix)

1. **Model** $m\langle \dot{v} \rangle = -kx$

$$S_2^{(1)}(x,v,t) = F_1\left(\frac{mv^2}{2} + \frac{kx^2}{2}, x\omega \sin(\omega t) + v\cos(\omega t)\right), \quad (1.18)$$

where $\omega = \sqrt{\dfrac{k}{m}}$ is the oscillator frequency; $F_1$ is some function of two variables.

**Remark 2**

Note that from (1.16) it follows that $\langle\langle Q_2 \rangle\rangle = 0$, and, according to (1.7), the function $H_2$ must be constant.

2. **Model** $m\langle \dot{v} \rangle = -kx - bv$

$$\xi_2(x,v) = \begin{cases} \ln|mv^2 + bvx + kx^2| - \dfrac{2b}{\sqrt{4mk-b^2}} \operatorname{arctg} \dfrac{2mv + bx}{x\sqrt{4mk-b^2}}, & b^2 < 4mk, \\[2ex] \ln|v + \omega x| + \dfrac{\omega x}{v + \omega x}, & b^2 = 4mk, \\[2ex] \ln|mv^2 + bvx + kx^2| - \dfrac{b}{\sqrt{b^2-4mk}} \ln\left|\dfrac{2mv + (b - \sqrt{b^2-4mk})x}{2mv + (b + \sqrt{b^2-4mk})x}\right|, & b^2 > 4mk, \end{cases} \quad (1.19)$$

$$\eta_2(x,v,t) = \begin{cases} \dfrac{v - p_1 x}{p_2 - p_1} e^{-p_2 t}, & b > 2m\omega, \\[2ex] xe^{\omega t}, & b = 2m\omega, \\[2ex] e^{\frac{b}{2m}t}\left[x\tilde{\omega}\sin(\tilde{\omega}t) + \left(v + \dfrac{b}{2m}x\right)\cos(\tilde{\omega}t)\right], & b < 2m\omega, \end{cases} \quad (1.20)$$

where $p_{1,2} = -\dfrac{b}{2m} \pm \tilde{\omega}$, $\tilde{\omega} = \dfrac{\sqrt{|b^2 - 4m^2\omega^2|}}{2m}$. The general solution is of the form:

$$S_2^{(2)}(x,v,t) = F_2(\xi_2(x,v), \eta_2(x,v,t)) - bt, \quad (1.21)$$

where $F_2$ is some function.

**Remark 3**

As the value of $Q_2 = -\dfrac{b}{m}$ (1.16) does not depend on the coordinate and velocity, it is possible to determine the average value $\langle\langle Q_2 \rangle\rangle$ without knowing the function $f_2(x,v,t)$. According to the expression (1.7), we obtain $\langle\langle Q_2 \rangle\rangle = -\dfrac{b}{m}$. From (1.7) it follows that the function $H_2$ is linear in $t$, that is $H_2(t) = \dfrac{b}{m}t + H_2(0)$.

3. **Model** $m\langle \dot{v} \rangle = -kx - \lambda v^2$

The general solution of the equation (1.4) is of the form:

$$S_2^{(3)}(x,v,t) = F_3\big(\xi_3(x,v), \eta_3(x,v,t)\big) + \frac{2\lambda}{m}x, \tag{1.22}$$

$$\xi_3(x,v) = e^{2\frac{\lambda}{m}x}\left(v^2 + \frac{k}{\lambda}x - \frac{mk}{2\lambda^2}\right), \tag{1.23}$$

$$\eta_3(x,v,t) = G(x,v) - t, \tag{1.24}$$

where $F_3$ is some function. The function $G(x,v)$ in (1.24) can be represented as follows:

$$G(x,v) = g\big(\xi_3(x,v), x\big), \tag{1.25}$$

$$g(C,x) = \int \frac{dx}{\sqrt{Ce^{-2\frac{\lambda}{m}x} - \frac{k}{\lambda}x + \frac{mk}{2\lambda^2}}}. \tag{1.26}$$

The constant value $C = const$ in the integral (1.26) acts as a parametric variable. After taking the integral (1.26) it is necessary to put the expression for the function $\xi_3(x,v)$ (1.23) instead of the parameter $C$ (argument of the function $g(C,x)$).

**Remark 4**

Let us consider the solution symmetry properties (1.22) with respect to the variable $v$. The function $\xi_3(x,v)$ depends only on $v^2$, therefore $\xi_3(x,-v) = \xi_3(x,v)$. The function $G(x,v)$ will contain the variable $v$ only after taking the integral $g(C,x)$ and setting into (1.25). Therefore, the function $G(x,v)$ has the property of symmetry $G(x,-v) = g\big(\xi_3(x,-v),x\big) = g\big(\xi_3(x,v),x\big) = G(x,v)$.

As a result, the solution (1.22) is symmetric in $v$, that is $S_2^{(3)}(x,v,t) = S_2^{(3)}(x,-v,t)$. A similar statement is correct for the function

$$f_2(x,v,t) = f_2(x,-v,t). \tag{1.27}$$

Then, calculating $\langle\langle Q_2 \rangle\rangle$, we obtain

$$\langle\langle Q_2\rangle\rangle(t) = \int_{-\infty}^{+\infty}\int_{-\infty}^{+\infty} f_2(x,v,t)Q_2(x,v,t)dxdv = -\frac{2\lambda}{m}\int_{-\infty}^{+\infty} dx \int_{-\infty}^{+\infty} f_2(x,v,t)vdv = 0, \qquad (1.28)$$

because by symmetry (1.27) $\int_{-\infty}^{+\infty} f_2(x,v,t)vdv = 0$. It follows from (1.28) that the Boltzmann function $H_2 = const$.

**§2 Numerical simulation**

Let us consider the $N$-particles system. For each particle at the initial moment of time, let us set the coordinate, the speed in accordance with some initial distribution. Integrating the motion equations (1.9) - (1.12) gives the trajectories along which the particles move.

Knowing the position, velocity and acceleration of the particles at each moment of time, it is possible to construct the distribution function $f_3(x,v,\dot{v},t)$. Knowing $f_3(x,v,\dot{v},t)$, by the formula (1.3), it is possible to determine the functions $f_2(x,v,t)$, $f_1(x,t)$ and $S_2(x,v,t)$, $S_1(x,t)$, as well as the average values (1.4) of the kinematic quantities $\langle v \rangle$ and $\langle \dot{v} \rangle$. According to the average values of kinematic quantities $\langle v \rangle$ and $\langle \dot{v} \rangle$, it is possible to determine the sources of dissipation $Q_1$, $Q_2$ (1.2) and their average values $\langle Q_1 \rangle$, $\langle\langle Q_2 \rangle\rangle$. Substituting the functions $S_1$, $Q_1$ and $S_2$, $Q_2$ into the equations (1.1) for the cases $n=1$ and $n=2$, respectively, we can verify the correctness of the equations (1.1). Knowing the functions $f_2(x,v,t)$ and $f_1(x,t)$, using the formulas (1.6), we can determine the generalized Boltzmann $H_1$ and $H_2$ functions. Substituting the obtained functions $H_1$, $\langle Q_1 \rangle$ and $H_2$, $\langle\langle Q_2 \rangle\rangle$ in the equations (1.5), we can verify the correctness of these equations.

On the other hand, we can set the functions $F_1, F_2, F_3$ in the expressions (1.18), (1.21), (1.22), respectively, and get the exact solutions $S_2^{(1)}, S_2^{(2)}, S_2^{(3)}$ for the equation (1.13). The exact solutions $S_2^{(1)}, S_2^{(2)}, S_2^{(3)}$ can be compared with solutions obtained by numerical integration of the motion equations (1.9) - (1.12).

In both cases, it is possible to estimate which of the equations (1.7) or (1.8) best describes dissipative systems.

Numerical integrating of the motion equations (1.9) - (1.12) was performed by the Verlet method [33]. Systems consisting of $N = 2^{23}$ particles were considered. The calculations were performed on the massively parallel GPU architecture using NVIDIA CUDA technology.

To assess the reliability of the results, statistical analysis of the data was carried out. The distributions of the quantities $\frac{dH_2}{dt}$ and $\langle\langle Q_2 \rangle\rangle$ in all cases of the simulation obey the normal law, which allows the use of a number of statistical criteria for testing hypotheses. Verification of this fact was carried out according to the Kolmogorov-Smirnov criterion based on the value $D$ [34-36]:

$$D = \sup |F_n(x) - F(x)|,$$

where $F_n(x)$ is a empirical distribution function, $F(x)$ is the normal distribution. Testing the equality of average values in two samples was carried out using Student's $t$-test [37, 38]:

$$t = \sqrt{N} \frac{\bar{X}_1 - \bar{X}_2}{\sqrt{\sigma_1^2 + \sigma_2^2}},$$

where $N$ is a sample size, $\bar{X}_1$ and $\bar{X}_2$ are sample averages, $\sigma_1^2$ and $\sigma_2^2$ are sample variances. When testing the hypotheses, the value of $p-value$ was measured [39–41]. In the case when the values $\frac{dH_2}{dt}$ and $\langle\langle Q_2 \rangle\rangle$ changed over time, a linear regression was constructed, the coefficients of which were calculated by the least squares method. The Pearson correlation coefficient $r$ and the determination coefficient $r^2$ were also calculated:

$$r_{X_1 X_2} = \frac{\text{cov}_{X_1 X_2}}{\sigma_{X_1} \sigma_{X_2}}.$$

## 2.1 Model $m\dot{v} = -kx$

In this system, the particles move in the harmonic oscillator potential and do not experience the dissipative forces action. Based on the motion equation (1.9), the value $\dot{v}$ is not explicitly dependent on $v$, therefore, according to Remark 2, the numerical values $\langle\langle Q_2 \rangle\rangle$ and $\frac{dH_2}{dt}$ should be close to zero. Fig. 1 shows the simulation results for the values $\langle\langle Q_2 \rangle\rangle$ and $\frac{dH_2}{dt}$. Fig. 1 (right) shows the numerical values of $\frac{dH_2}{dt}$ and $\langle\langle Q_2 \rangle\rangle$ at different points in time. Fig. 1 (left) shows the points position with the coordinates $\left( \langle\langle Q_2 \rangle\rangle, \frac{dH_2}{dt} \right)$ at different points in time.

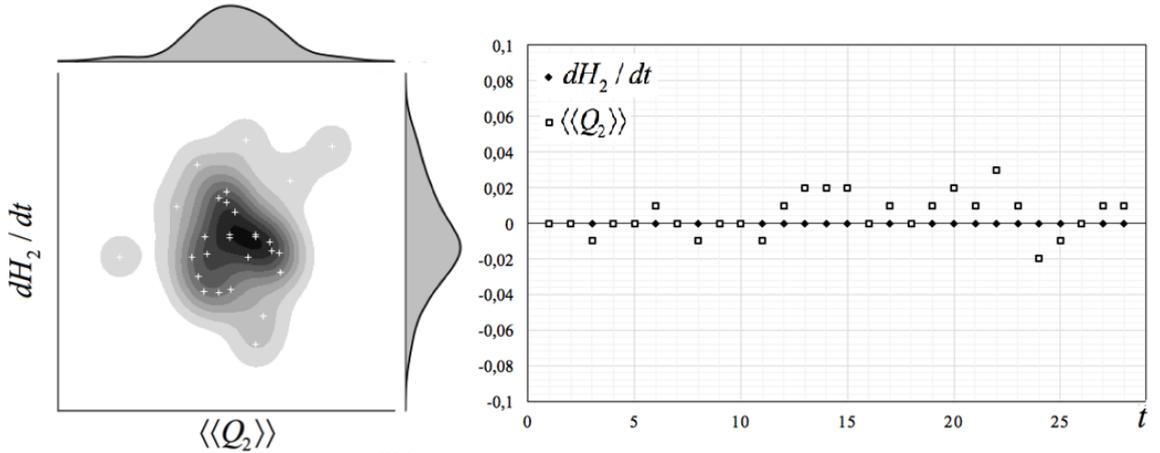

Fig. 1 Distribution of $\frac{dH_2}{dt}$ and $\langle\langle Q_2 \rangle\rangle$ in the case $m\dot{v} = -kx$

Fig. 1 shows the agreement of the simulation results with the theoretical predictions (see Remark 2). Both quantities $\frac{dH_2}{dt}$ and $\langle\langle Q_2 \rangle\rangle$ have numerical values close to zero. The result reliability is confirmed by the Student's t-test ($t = 1.89$ and $p_{value}^{(s)} = 0.06$). The values of $\frac{dH_2}{dt}$ and $\langle\langle Q_2 \rangle\rangle$ have normal distributions (see Fig. 1 left), as evidenced by the Kolmogorov-Smirnov criterion ($D = 0.147$, $p_{value}^{(KS)} = 0.54$).

From expressions (1.1) and (1.16) it follows that $\frac{d_2 S_2}{dt}$ and $Q_2$ do not depend on time and are equal to zero. Fig. 2 shows the distribution of $\frac{d_2 S_2}{dt}$ and $Q_2$ in the plane $XOV$, obtained by numerical simulation. Fig. 2 shows good agreement between the theoretical prediction and numerical simulation.

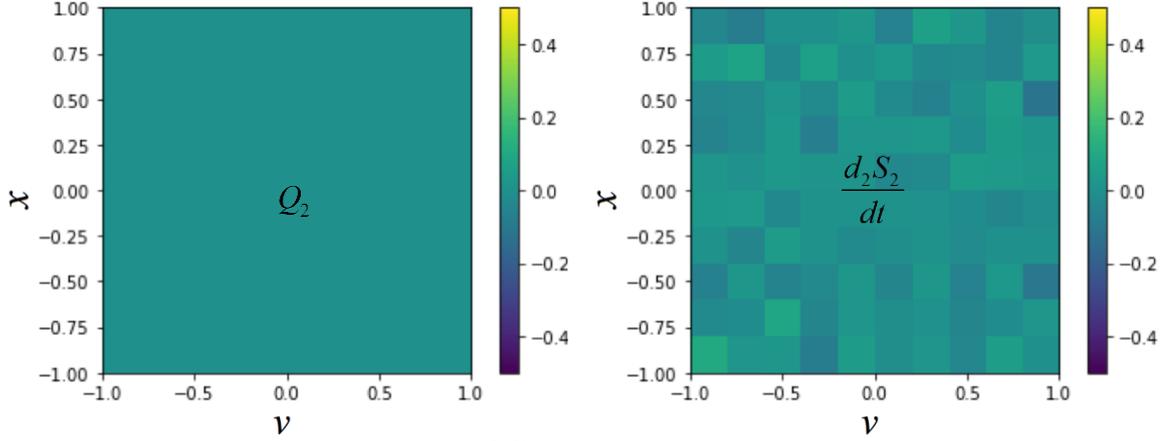

Fig. 2 Distribution of $Q_2$ and $\frac{d_2 S_2}{dt}$ in the case $m\dot{v} = -kx$

As a result, in the case $m\dot{v} = -kx$, there are no dissipations ($Q_2 = 0$) in the system, and $H_2 = const$. Such a system is described by the classical Vlasov equation (i.1), and the corresponding equation (1.8).

**2.2 Model $m\dot{v} = -kx - bv$**

The motion equation (1.10) contains the viscous friction force proportional to velocity $-bv$. In this case, there are the dissipations sources $Q_2 = -\frac{b}{m}$ (1.16), and, according to Remark 3, $\langle\langle Q_2 \rangle\rangle = -\frac{b}{m}$. Calculations were made for different values of the coefficient $\frac{b}{m}$. Figs. 3-6 show the numerical simulation results for the values $\frac{b}{m} = \{0.5, 0.8, 1.0\}$. $D$ and $p_{value}^{(KS)}$ have values $\{0.1, 0.15, 0.1\}$ and $\{0.94, 0.53, 0.92\}$, respectively. The obtained values of $D$ and $p_{value}^{(KS)}$ allow us to use the Student's t-test. The values $t = \{0.9, 0.3, -1.7\}$ and $p_{value}^{(S)} = \{0.38, 0.77, 0.09\}$ allow us to accept the hypothesis of equality of the averages value of $\frac{dH_2}{dt}$ and $\langle\langle Q_2 \rangle\rangle$. In Figs. 3-5, it is clear that the numerical values of $\frac{dH_2}{dt}$ and $\langle\langle Q_2 \rangle\rangle$ are close to the theoretical value $-\frac{b}{m}$ (see Remark 3).

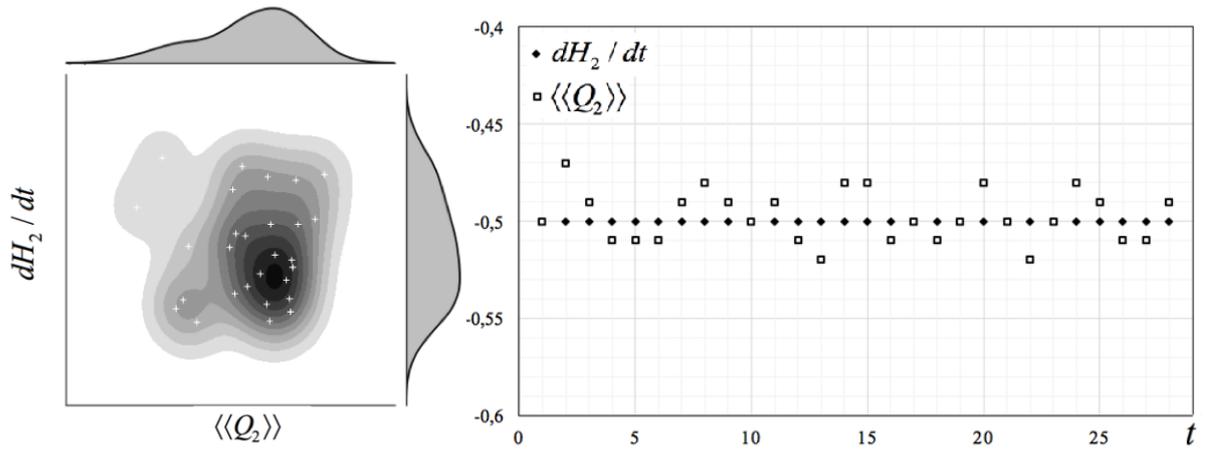

Fig. 3 Distribution of $\dfrac{dH_2}{dt}$ and $\langle\langle Q_2 \rangle\rangle$ in the case $\dot{v} = -\dfrac{k}{m}x - 0.5v$

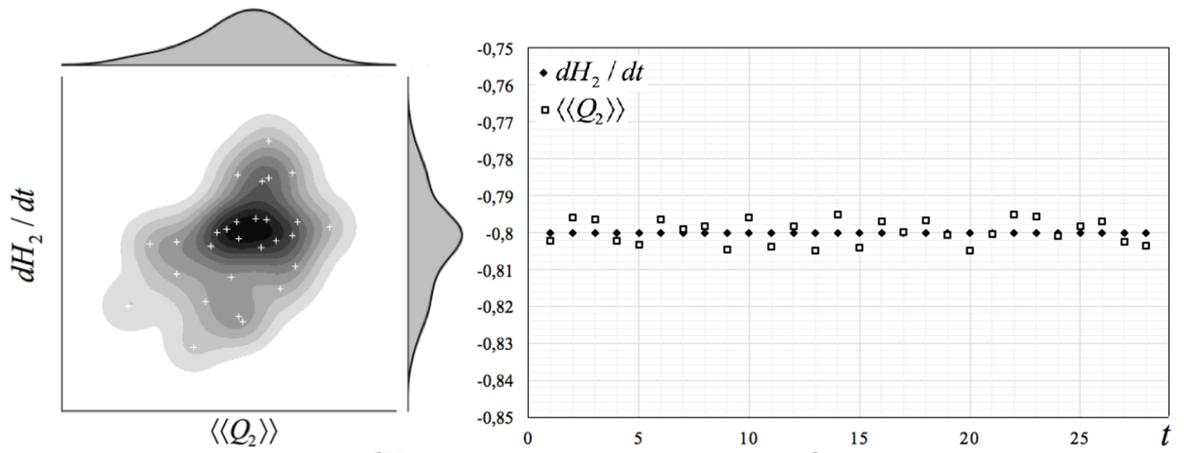

Fig. 4 Distribution of $\dfrac{dH_2}{dt}$ and $\langle\langle Q_2 \rangle\rangle$ in the case $\dot{v} = -\dfrac{k}{m}x - 0.8v$

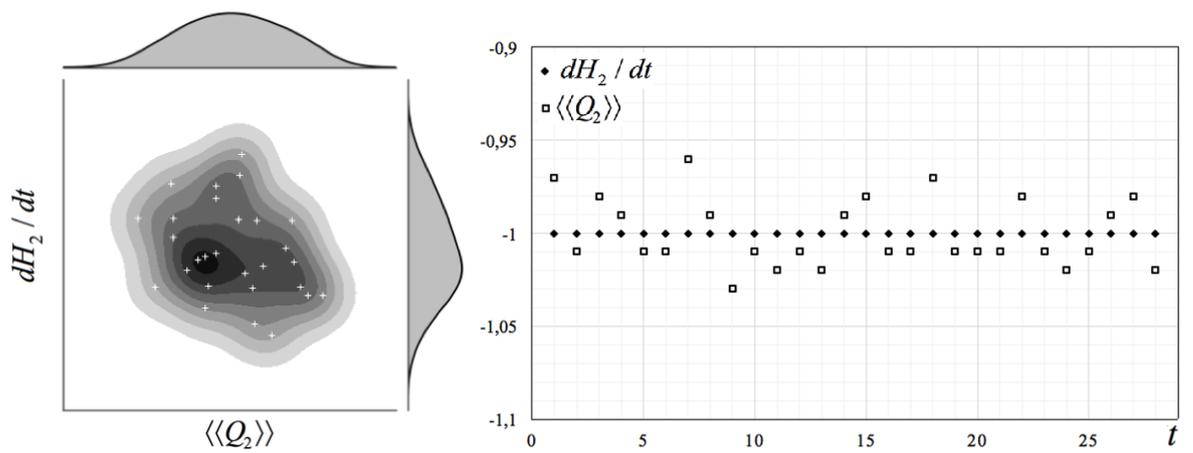

Fig. 5 Distribution of $\dfrac{dH_2}{dt}$ and $\langle\langle Q_2 \rangle\rangle$ in the case $\dot{v} = -\dfrac{k}{m}x - v$

The values of $\frac{d_2 S_2}{dt}$ and $Q_2$ do not depend on time and have a constant value $-\frac{b}{m}$. Fig. 6 shows the distributions of $\frac{d_2 S_2}{dt}$ and $Q_2$, found numerically at $\frac{b}{m} = 0.5$. In Fig. 6, a good agreement of the theoretical and calculated data is seen.

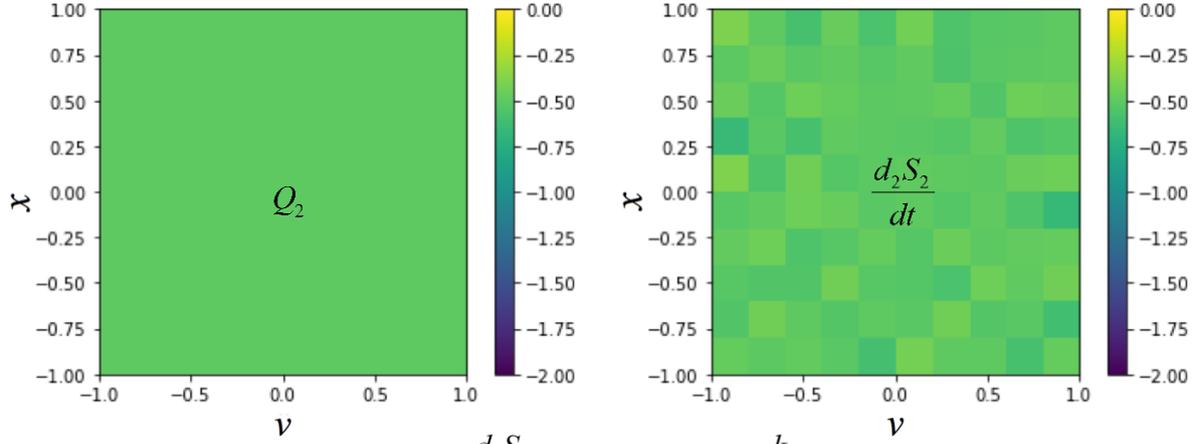

Fig. 6 Distribution of $Q_2$ and $\frac{d_2 S_2}{dt}$ in the case $\dot{v} = -\frac{k}{m}x - 0.5v$

Thus, it is impossible to describe the dissipative system $m\dot{v} = -kx - bv$ by the classical Vlasov equation (i.1) and the equation (1.8) (since the right-hand side $Q_2 \neq 0$ and $\langle\langle Q_2 \rangle\rangle \neq 0$). On the other hand, from the theoretical predictions (Remark 3) and the results of numerical simulation (see Fig. 3-6), it is clear that the dissipative system $m\dot{v} = -kx - bv$ is well described by the modified (original) Vlasov equation (i.4), (1.13) and the equation (1.7).

**2.3 Model** $m\dot{v} = -kx - \lambda v^2$

In this system, there is the viscous friction force $-\lambda v^2$, where $\frac{\lambda}{m} = 0.5$. Despite the presence of the dissipative sources $Q_2 = \frac{\lambda}{m}v$, their average value is zero, that is $\langle\langle Q_2 \rangle\rangle = 0$ (see Remark 4 (1.28)). Fig. 7 shows that the numerical values of $\frac{dH_2}{dt}$ and $\langle\langle Q_2 \rangle\rangle$ are close to zero. The value $D = 0.12$ and $p_{value}^{(KS)} = 0.8$, therefore, we can use the Student's t-criterion. The values $t = -0.62$ and $p_{value}^{(S)} = 0.53$ allow us to accept the hypothesis of equality of $\frac{dH_2}{dt}$ and $\langle\langle Q_2 \rangle\rangle$.

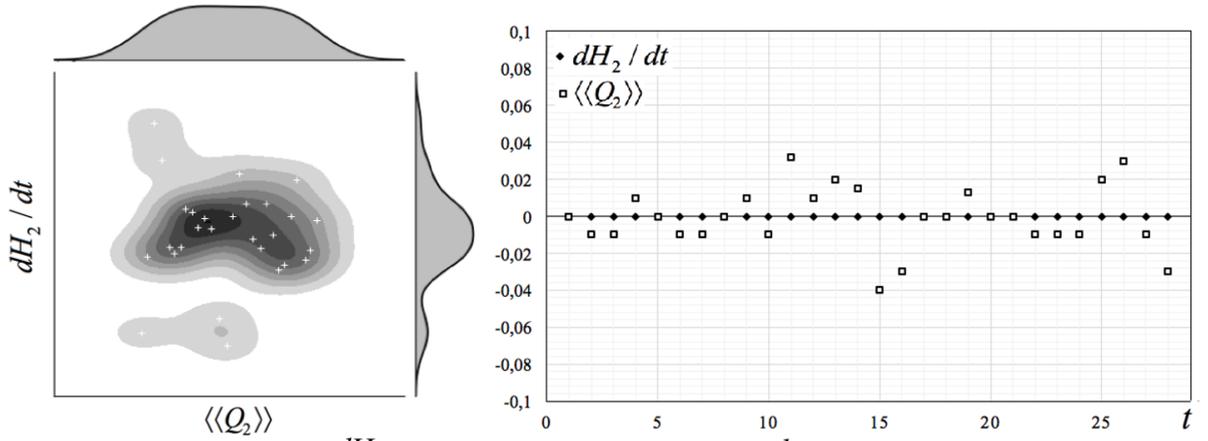

Fig. 7 Distribution of $\dfrac{dH_2}{dt}$ and $\langle\langle Q_2\rangle\rangle$ in the case $\dot{v}=-\dfrac{k}{m}x-0.5v^2$

According to the theoretical predictions, the quantities $\dfrac{d_2 S_2}{dt}$ and $Q_2$ do not depend on time, but depend on the velocity $v$. Fig. 8 shows the numerical distribution of $\dfrac{d_2 S_2}{dt}$ and $Q_2$ in the plane $XOV$ at $\dfrac{\lambda}{m}=0.5$.

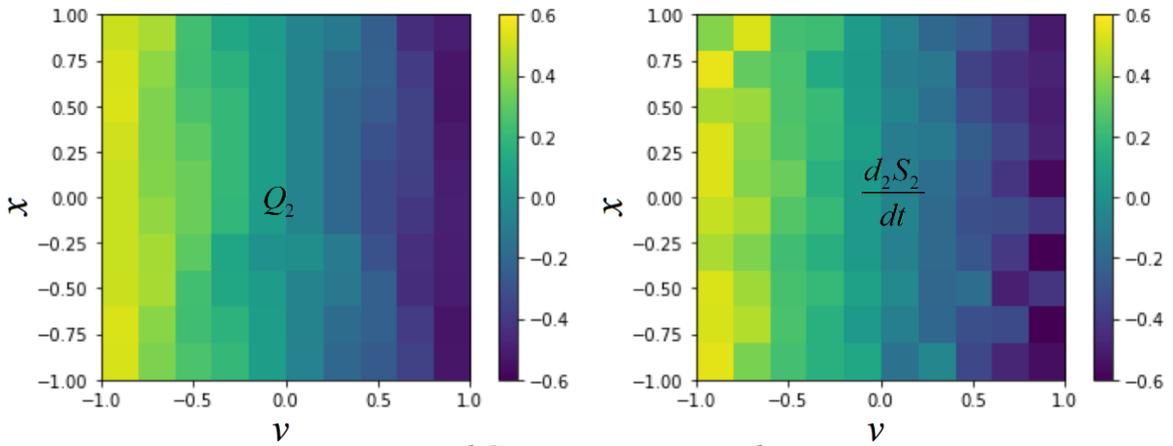

Fig. 8 Distribution of $Q_2$ and $\dfrac{d_2 S_2}{dt}$ in the case $\dot{v}=-\dfrac{k}{m}x-0.5v^2$

According to Remark 1, the description of the system $m\dot{v}=-kx-\lambda v^2$ requires the use of the modified (original) Vlasov equation (i.4), (1.13), although the Boltzmann $H_2$-function satisfies equation (1.8).

### 2.4 Model $m\dot{v}=-kx-A\cos(\gamma v)$

In this case, the system has the dissipation sources $Q_2 = A\gamma \sin(\gamma v)$ (1.16). An analytical expression for the dissipation sources $\langle\langle Q_2\rangle\rangle$ is difficult to obtain. However, as in the previous cases, we can check the satisfiability of the equation (1.7) for the Boltzmann $H_2$-function and the satisfiability of the modified (original) Vlasov equation (1.13).

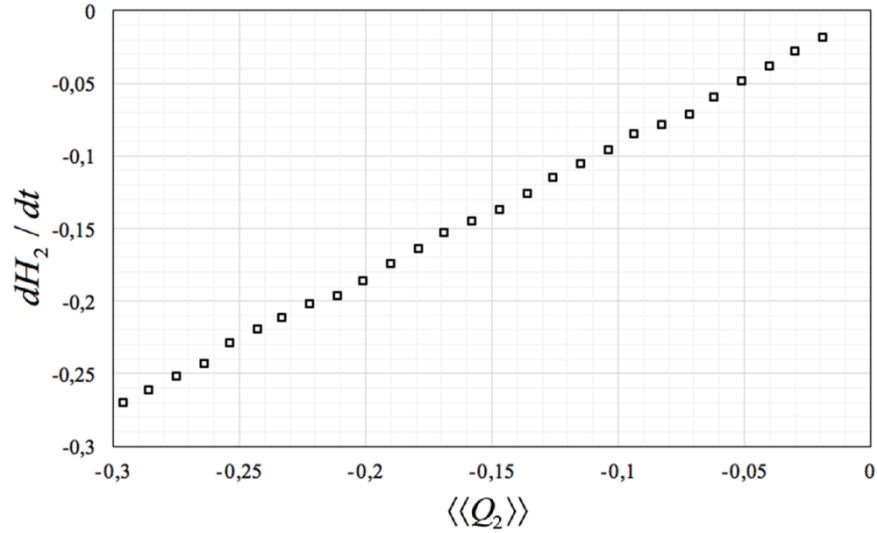

Fig. 9 Dependency graph of $\frac{dH_2}{dt}$ from $\langle\langle Q_2 \rangle\rangle$ in the case $m\dot{v} = -kx - A\cos(\gamma v)$

Fig. 9 shows the numerical simulation results for $\frac{dH_2}{dt}$ and $\langle\langle Q_2 \rangle\rangle$. In fig. 9 it is seen that the values $\frac{dH_2}{dt}$ and $\langle\langle Q_2 \rangle\rangle$ change over time. The data shown in Fig. 9 correspond to the linear regression $\frac{dH_2}{dt} = \alpha \langle\langle Q_2 \rangle\rangle$ with the linear regression coefficient $\alpha = 0.92 \pm 0.14$, the correlation coefficient $r = 0,998$, and the determination coefficient $r^2 = 0.996$.

According to the theory, $\frac{d_2 S_2}{dt}$ and $Q_2$ do not depend on time, but depend on the velocity $v$ (1.16). Fig. 10 shows the numerical distribution of $\frac{d_2 S_2}{dt}$ and $Q_2$ at some point in time $t$.

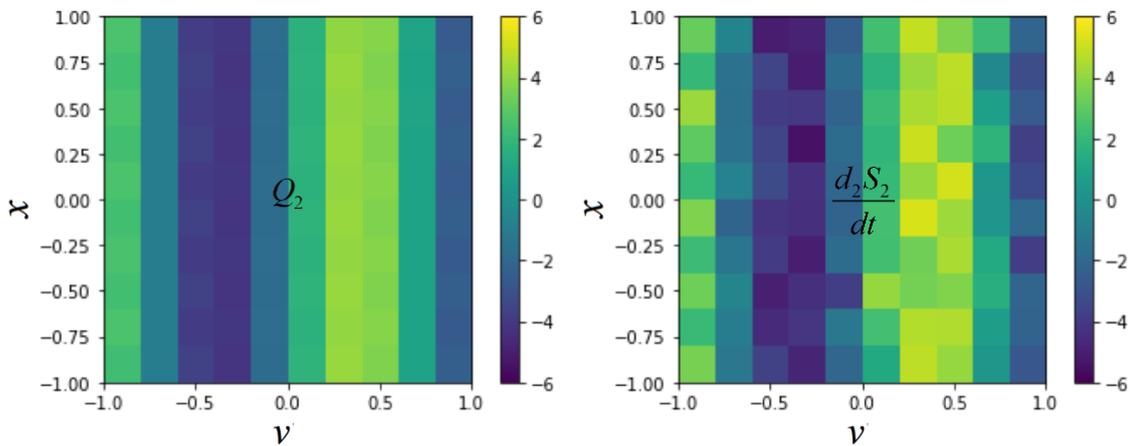

Fig. 10 Distribution of $Q_2$ and $\frac{d_2 S_2}{dt}$ in the case $m\dot{v} = -kx - A\cos(\gamma v)$

Thus, the numerical experiment confirms the correctness of using the modified (original) Vlasov equation (i.4), (1.13) and the equation (1.7) in the description of the dissipative process $m\dot{v} = -kx - A\cos(\gamma v)$.

## §3 Discussion

Let us discuss the numerical simulation results (see §2). In all four considered systems (1.9) - (1.12), the numerical simulation results are consistent with good accuracy with the theoretical predictions from §1.

The system (1.9) corresponds to the usual harmonic oscillator, for which the total energy remains constant. Therefore, it was natural to expect the absence of the dissipation sources $Q_2 = 0$ and $\langle\langle Q_2 \rangle\rangle = 0$, and the classical Vlasov equation (i.8) implementation and the equation (1.8). Hence, the Boltzmann $H_2$-function is constant.

Systems (1.10) - (1.12) are dissipative with a nonzero value of $Q_2 \neq 0$. In accordance with the theoretical predictions and the numerical experiment results, it is necessary to use the modified (original) Vlasov equation (1.13) to describe the systems (1.10) - (1.12). Using the classical Vlasov equation (i.4) for the systems (1.10) - (1.12) will give an incorrect result.

For the dissipative system (1.11) with the friction force $-\lambda v^2$, the dissipation sources are of the form $Q_2 = -2\frac{\lambda}{m}v$. The average value of dissipation sources is zero, that is $\langle\langle Q_2 \rangle\rangle = 0$. This result was predicted theoretically (see Remark 3) and confirmed by the numerical simulation (see 2.3). For a correct description of the system (1.11), it is necessary to use the modified (original) Vlasov equation (1.13). Despite the presence of the dissipation sources $Q_2 \neq 0$, the Boltzmann $H_2$-function is a constant since $\langle\langle Q_2 \rangle\rangle = 0$.

Dissipative systems (1.10) and (1.12) have nonzero the dissipations sources $Q_2 \neq 0$ and nonzero the average values $\langle\langle Q_2 \rangle\rangle \neq 0$. The both systems (1.10) and (1.12) are described by the modified (original) Vlasov equation (1.13) and the corresponding equation for the Boltzmann $H_2$-function (1.8).

In Figs. 1, 3, 4, 5, 7 (left) one can see that the systems (1.9) - (1.11) have stationary distributions of average values of the dissipation sources $\langle\langle Q_2 \rangle\rangle$ (points with the coordinates $\left(\langle\langle Q_2 \rangle\rangle, \frac{dH_2}{dt}\right)$ are localized in a neighborhood of some center). For the systems (1.9) - (1.10), the Boltzmann $H_2$-function in the general case has the form $H_2(t) = c_1 t + c_2$, where $c_1 = -\langle\langle Q_2 \rangle\rangle$, $c_2 = H_2(0)$ are constant values.

The system (1.12) has the non-stationary distribution of average sources of dissipation $\langle\langle Q_2 \rangle\rangle$ (see Fig. 9). As a result, the Boltzmann $H_2$-function has the non-linear time dependence.

## Conclusion

From the results obtained in the paper, it follows that when considering dissipative systems, it is necessary to use the modified (original) Vlasov equation (i.4). The use of the classical Vlasov equation (i.1) in the dissipative systems description ($Q_2 \neq 0$) is incorrect. This fact explains the attempts to change the Vlasov (Enskog-Vlasov) equation undertaken in various papers.

Instead of introducing semi-phenomenological expressions for the right-hand sides of the Vlasov equation, we offer the natural form of the equation (modified Vlasov equation (i.4) [32]), obtained from first principles by A. A. Vlasov himself.

The equation (i.4) in the general case can be used for dissipative and conservative systems.

In this paper, we examined the numerical simulation of a system described only by the distribution functions $f_1(\vec{r},t), f_2(\vec{r},\vec{v},t)$. In the general case, the physical system is described by an infinite set of functions $f_1(\vec{r},t), f_2(\vec{r},\vec{v},t), f_3(\vec{r},\vec{v},\dot{\vec{v}},t),...$ satisfying the chain of Vlasov equations (1.1) [32].

The method described in this paper can model dissipative systems, whose dissipations are due to higher kinematic quantities $Q_n$, $n > 2$, which corresponds, for example, to the presence of radiation in the system. Such an approach makes it possible to consider the problems of the plasma physics, the accelerator physics, the quantum mechanics, the astrophysics and the problems of controlled thermonuclear fusion in a new way, based on first principles, without involving phenomenological considerations.

**Acknowledgements**

This work was supported by the RFBR No. 18-29-10014.

**Appendix**

1. **Model** $m\langle\dot{v}\rangle = -kx$

From the equations (1.17) it follows that

$$\frac{dx}{mv} = -\frac{dv}{kx}, \quad C_1 = \frac{mv^2}{2} + \frac{kx^2}{2}. \tag{A.1}$$

Note that the first integral (A.1) corresponds to the total energy of the harmonic oscillator. We obtain the second integral:

$$\frac{dx}{dt} = v, \quad \frac{dv}{dt} = \langle\dot{v}\rangle = -\frac{k}{m}x, \quad \frac{d^2x}{dt^2} + \omega^2 x = 0, \tag{A.2}$$

where $\omega = \sqrt{\dfrac{k}{m}}$ is the frequency of the oscillator. From the expressions (A.2), we obtain

$$x = A\cos(\omega t) + B\sin(\omega t), \quad v = -A\omega\sin(\omega t) + B\omega\cos(\omega t),$$
$$C_2 = x\omega\sin(\omega t) + v\cos(\omega t). \tag{A.3}$$

2. **Model** $m\langle\dot{v}\rangle = -kx - bv$

The characteristic equations (1.17) will be of the form:

$$\frac{dt}{1} = \frac{dx}{v} = -\frac{mdv}{kx+bv} = \frac{mdS_2}{b}, \tag{A.4}$$

hence, the first integral has the form

$$bdt = mdS_2, \quad C_1 = mS_2 - bt. \tag{A.5}$$

We obtain the second integral

$$\frac{dx}{v} = -\frac{mdv}{kx+bv}, \quad -\frac{dv}{dx} = \frac{kx+bv}{mv}. \tag{A.6}$$

Let us determine the function $v(x) = y(x)x$, where $y(x)$ is some unknown function. Then from (A.6), we obtain

$$-\frac{dv}{dx} = -y - xy' = \frac{kx+bxy}{mxy} = \frac{k+by}{my}, \quad -xy' = \frac{k+by+my^2}{my},$$

$$\frac{mydy}{k+by+my^2} = -\frac{dx}{x}. \tag{A.7}$$

Integrating the equation (A.7), we obtain

$$\int \frac{mydy}{my^2+by+k} = \int \frac{\bar{y}d\bar{y}}{\bar{y}^2+\mu} - \frac{b}{2\sqrt{m}}\int \frac{d\bar{y}}{\bar{y}^2+\mu} = \frac{1}{2}\ln\left|\bar{y}^2+\mu\right| - \frac{b}{2\sqrt{m}}\int \frac{d\bar{y}}{\bar{y}^2+\mu}. \tag{A.8}$$

where $\bar{y} = \sqrt{m}y + \frac{b}{2\sqrt{m}}$, $\mu = k - \frac{b^2}{4m}$. Depending on the $\mu$, the second integral in (A.8) may have different representations:

$$\int \frac{d\bar{y}}{\bar{y}^2+\mu} = \begin{cases} -\dfrac{1}{\bar{y}}, & npu \ 4mk = b^2, \ \mu = 0, \\ \dfrac{1}{\sqrt{\mu}}\operatorname{arctg}\dfrac{\bar{y}}{\sqrt{\mu}}, & npu \ 4mk > b^2, \ \mu > 0, \\ \dfrac{1}{2\sqrt{|\mu|}}\ln\left|\dfrac{\bar{y}-\sqrt{|\mu|}}{\bar{y}+\sqrt{|\mu|}}\right|, & npu \ 4mk < b^2, \ \mu < 0. \end{cases} \tag{A.9}$$

For the second integral there are three possible representations. Let us write each of them. In the first variant (A.9) with $4mk = b^2$ from (A.7) - (A.9) it follows

$$const - \ln|x| = \frac{1}{2}\ln|\bar{y}^2| + \frac{\sqrt{k}}{\bar{y}}, \quad const = \ln(\bar{y}x) + \frac{\sqrt{k}}{\bar{y}},$$

$$C_2^{(1)} = \ln\left[x(y+\omega)\right] + \frac{\omega}{y+\omega},$$

$$C_2^{(1)} = \ln(v+\omega x) + \frac{\omega x}{v+\omega x}. \tag{A.10}$$

In the second variant (A.9) with $4mk > b^2$, we obtain

$$const - \ln|x| = \frac{1}{2}\ln|\bar{y}^2 + \mu| - \frac{b}{2\sqrt{m\mu}} \operatorname{arctg} \frac{\bar{y}}{\sqrt{\mu}},$$

$$C_2^{(2)} = \ln\left[x^2\left(my^2 + by + k\right)\right] - \frac{2b}{\sqrt{4mk - b^2}} \operatorname{arctg} \frac{2my + b}{\sqrt{4mk - b^2}},$$

$$C_2^{(2)} = \ln\left(mv^2 + bvx + kx^2\right) - \frac{2b}{\sqrt{4mk - b^2}} \operatorname{arctg} \frac{2mv + bx}{x\sqrt{4mk - b^2}}. \tag{A.11}$$

In the third variant (A.9) with $4mk < b^2$, respectively

$$const - \ln|x| = \frac{1}{2}\ln|\bar{y}^2 + \mu| - \frac{b}{4\sqrt{m|\mu|}} \ln\left|\frac{\bar{y} - \sqrt{|\mu|}}{\bar{y} + \sqrt{|\mu|}}\right|,$$

$$C_2^{(3)} = \ln\left[x^2\left|my^2 + by + k\right|\right] - \frac{b}{\sqrt{b^2 - 4mk}} \ln\left|\frac{2my + b - \sqrt{b^2 - 4mk}}{2my + b + \sqrt{b^2 - 4mk}}\right|,$$

$$C_2^{(3)} = \ln\left|mv^2 + bvx + kx^2\right| - \frac{b}{\sqrt{b^2 - 4mk}} \ln\left|\frac{2mv + \left(b - \sqrt{b^2 - 4mk}\right)x}{2mv + \left(b + \sqrt{b^2 - 4mk}\right)x}\right|. \tag{A.12}$$

Let us find the third integral from the equations (A.4), we obtain

$$\frac{dx}{dt} = v, \quad \frac{dv}{dt} = \langle \dot{v} \rangle = -\frac{k}{m}x - \frac{b}{m}v,$$

$$\frac{d^2x}{dt^2} + \vartheta \frac{dx}{dt} + \omega^2 x = 0, \tag{A.13}$$

$$p^2 + \vartheta p + \omega^2 = 0, \quad p_{1,2} = \frac{-\vartheta \pm \sqrt{\vartheta^2 - 4\omega^2}}{2},$$

where $\vartheta = \frac{b}{m}$. There are two possible types of solutions of the equation (A.13), corresponding to real and complex roots of $p_{1,2}$. Let us consider each case. We start with the real values of $p_{1,2}$, that is $\vartheta \geq 2\omega$:

$$x = Ae^{p_1 t} + Be^{p_2 t}, \quad v = Ap_1 e^{p_1 t} + Bp_2 e^{p_2 t}, \quad v - p_1 x = Be^{p_2 t}(p_2 - p_1),$$

$$C_3^{(1)} = \frac{v - p_1 x}{p_2 - p_1} e^{-p_2 t}. \tag{A.14}$$

If $\vartheta = 2\omega$, then $p_1 = p_2 = p = -\frac{\vartheta}{2}$, and the expression (A.14) is of the form

$$x = Ae^{pt}, \qquad C_3^{(1)} = xe^{-pt}. \tag{A.15}$$

In a similar manner, for the complex values of $p_{1,2}$ ($\vartheta < 2\omega$) we obtain:

$$xe^{-pt} = A\cos(\tilde{\omega}t) + B\sin(\tilde{\omega}t), \quad \tilde{\omega} = \sqrt{\omega^2 - p^2},$$

$$e^{-pt}\frac{v - px}{\tilde{\omega}} = -A\sin(\tilde{\omega}t) + B\cos(\tilde{\omega}t),$$

$$C_3^{(2)} = e^{-pt}\left[x\tilde{\omega}\sin(\tilde{\omega}t) + (v - px)\cos(\tilde{\omega}t)\right]. \tag{A.16}$$

3. **Model** $m\langle\dot{v}\rangle = -kx - \lambda v^2$

The characteristic equation (1.17) is as follows

$$\frac{dt}{1} = \frac{dx}{v} = -\frac{mdv}{kx + \lambda v^2} = \frac{mdS_2}{2\lambda v}. \tag{A.17}$$

The first integral will be of the form:

$$\frac{2\lambda}{m}dx = dS_2, \quad C_1 = S_2 - \frac{2\lambda}{m}x. \tag{A.18}$$

The second integral will be written as follows

$$dx = -\frac{mvdv}{kx + \lambda v^2} = -\frac{m}{2}\frac{dv^2}{kx + \lambda v^2} = -\frac{m}{2}\frac{dy}{kx + \lambda y}, \tag{A.19}$$

where the designation $y = v^2$ is introduced. Further, from (A.19) we obtain a non-uniform linear differential equation

$$\frac{dy}{dx} + 2\frac{\lambda}{m}y = -2\frac{kx}{m} = -2\omega^2 x. \tag{A.20}$$

The solution of the equation (A.20) is of the form

$$y_{o.o} = Ae^{-2\frac{\lambda}{m}x}, \quad y_{ч.н.} = -\frac{k}{\lambda}x + \frac{mk}{2\lambda^2}, \quad y_{о.н.} = y_{o.o} + y_{ч.н.} \tag{A.21}$$

The particular solution of the inhomogeneous equation $y_{ч.н.}$ in (A.21) is obtained by varying an arbitrary constant. From (A.21) it follows that the second integral will be as follows:

$$C_2 = e^{2\frac{\lambda}{m}x}\left(v^2 + \frac{k}{\lambda}x - \frac{mk}{2\lambda^2}\right). \tag{A.22}$$

We find the third integral. From (A.17) we obtain

$$m\frac{d^2x}{dt^2} + \lambda\left(\frac{dx}{dt}\right)^2 = -kx. \tag{A.23}$$

Let us introduce the designation $p(x) = \frac{dx}{dt}$, then $\frac{dp}{dt} = p'p$, and the equation (A.23) will be of the form:

$$mp'p + \lambda p^2 = -kx,$$
$$m(p^2)' + 2\lambda p^2 = -2kx,$$
$$my' + 2\lambda y = -2kx, \qquad (A.24)$$

where $y = p^2$. The equation (A.24) coincides with the equation (A.20), therefore we can write

$$\frac{dx}{dt} = p = \sqrt{C_2 e^{-2\frac{\lambda}{m}x} - \frac{k}{\lambda}x + \frac{mk}{2\lambda^2}},$$

$$C_3 = \int \frac{dx}{\sqrt{C_2 e^{-2\frac{\lambda}{m}x} - \frac{k}{\lambda}x + \frac{mk}{2\lambda^2}}} - t = g(C_2(x,v), x) - t = G(x,v) - t. \qquad (A.25)$$

**References**

1. Vlasov A.A., Many-Particle Theory and Its Application to Plasma, New York, Gordon and Breach, 1961, ISBN 0-677-20330-6; ISBN 978-0-677-20330-0
2. Vlasov A.A., Statisticheskie funkcii raspredelenija, Moscow, Nauka, 1966, 356 p.
3. Perepelkin E.E., Sadovnikov B.I., Inozemtseva N.G., The properties of the first equation of the Vlasov chain of equations, J. Stat. Mech. (2015) P05019
4. Boris Atenas, Sergio Curilef, Dynamics and thermodynamics of systems with long-range dipole-type interactions, Physical Review E 95, 022110 (2017)
5. Massimiliano Giona, Space-time-modulated stochastic processes, Physical Review E 96, 042132 (2017)
6. Pankaj Kumar and Bruce N. Miller, Thermodynamics of a one-dimensional self-gravitating gas with periodic boundary conditions, Physical Review E 95, 022116 (2017)
7. M. E. Carrington, St. Mrowczynski, B. Schenke, Momentum broadening in unstable quark-gluon plasma, Physical Review C 95, 024906 (2017)
8. R. Haenel, M. Schulz-Weiling, J. Sous, H. Sadeghi, M. Aghigh, L. Melo, J. S. Keller, E. R. Grant, Arrested relaxation in an isolated molecular ultracold plasma, Physical Review A 96, 023613 (2017)
9. Kentaro Hara, Ido Barth, Erez Kaminski, I. Y. Dodin, N. J. Fisch, Kinetic simulations of ladder climbing by electron plasma waves, Physical Review E 95, 053212 (2017)
10. M. Horky, W. J. Miloch, V. A. Delong, Numerical heating of electrons in particle-in-cell simulations of fully magnetized plasmas, Physical Review E 95, 043302 (2017)
11. N. Ratan, N. J. Sircombe, L. Ceurvorst,1 J. Sadler, M. F. Kasim, J. Holloway, M. C. Levy, R. Trines, R. Bingham, and P. A. Norreys, Dense plasma heating by crossing relativistic electron beams, Physical Review E 95, 013211 (2017)
12. Shetty, D. V., Botvina, A. S., Yennello, S. J., Souliotis, G. A., Bell, E., & Keksis, A. (2005). Fragment yield distribution and the influence of neutron composition and excitation energy in multifragmentation reactions. Physical Review C, 71(2).
13. Zheng, H., Burrello, S., Colonna, M., Lacroix, D., & Scamps, G. (2018). Connecting the nuclear equation of state to the interplay between fusion and quasifission processes in low-energy nuclear reactions. Physical Review C, 98(2).
14. Pierroutsakou, D., Martin, B., Agodi, C., Alba, R., Baran, V., Boiano, A., … Signorini, C. (2009). Dynamical dipole mode in fusion reactions at 16 MeV/nucleon and beam energy dependence. Physical Review C, 80(2).
15. M. Kopp, K. Vattis, C. Skordis, Solving the Vlasov equation in two spatial dimensions with the Schrödinger method, Physical Review D 96, 123532 (2017)

16. S. Bergström, R. Catena, A. Chiappo, J. Conrad, B. Eurenius, M. Eriksson, M. Högberg, S. Larsson, E. Olsson, A. Unger, R. Wadman, J-factors for self-interacting dark matter in 20 dwarf spheroidal galaxies, Physical Review D 98, 043017 (2018)
17. L. Gabriel Gomez, J. A. Rueda, Dark matter dynamical friction versus gravitational wave emission in the evolution of compact-star binaries, Physical Review D 96, 063001 (2017)
18. Derek Inman, Hao-Ran Yu, Hong-Ming Zhu, J. D. Emberson, Ue-Li Pen, Tong-Jie Zhang, Shuo Yuan, Xuelei Chen, Zhi-Zhong Xing, Simulating the cold dark matter-neutrino dipole with TianNu, Physical Review D 95, 083518 (2017)
19. Giovanni Manfredi, Jean-Louis Rouet, Bruce Miller, Gabriel Chardin, Cosmological structure formation with negative mass, Physical Review D 98, 023514 (2018)
20. Jan Veltmaat, Jens C. Niemeyer, Bodo Schwabe, Formation and structure of ultralight bosonic dark matter halos, Physical Review D 98, 043509 (2018)
21. S. V. Batalov, A. G. Shagalov, Autoresonant excitation of Bose-Einstein condensates, Physical Review E 97, 032210 (2018)
22. Wojciech Florkowski, Ewa Maksymiuk, Radoslaw Ryblewski, Anisotropic-hydrodynamics approach to a quark-gluon fluid mixture, Physical Review C 97, 014904 (2018)
23. M. Stephanov1 and Y. Yin, Hydrodynamics with parametric slowing down and fluctuations near the critical point, Physical Review D 98, 036006 (2018)
24. E. Camporealea, G.L. Delzanno, B.K. Bergen, J.D. Moulton, On the velocity space discretization for the Vlasov–Poisson system: Comparison between implicit Hermite spectral and Particle-in-Cell methods, Computer Physics Communications (2016)
25. M. R. Dorr, P. Colella, M. A. Dorf, D. Ghosh, J. Hittinger, P. O. Schwartz, 6. High-order Discretization of a Gyrokinetic Vlasov Model in Edge Plasma Geometry, Journal of Computational Physics (2018)
26. E. Fijalkow, A numerical solution to the Vlasov equation, Computer Physics Communications (1999)
27. F. Filbet, E. Sonnendrucker, P. Bertrandz, Conservative Numerical Schemes for the Vlasov Equation, Journal of Computational Physics (2001)
28. E. Sonnendrucker, J. Roche, P. Bertrand, and A. Ghizzoy, The Semi-Lagrangian Method for the Numerical Resolution of the Vlasov Equation, Journal of Computational Physics (1999)
29. F. Valentini, P. Travnicek, F. Califano, P. Hellinger, A. Mangeney, A hybrid-Vlasov model based on the current advance method for the simulation of collisionless magnetized plasma, Journal of Computational Physics 225 (2007) 753–770
30. M. Grmela, Kinetic Equation Approach to Phase Transitions , Journal of Statistical Physics, Vol. 3, No. 3, 1971
31. E. S. Benilov, M. S. Benilov, Energy conservation and H theorem for the Enskog-Vlasov equation, Physical Review E 97, 062115 (2018)
32. E. E. Perepelkin, B. I. Sadovnikov, N. G. Inozemtseva, The new modified Vlasov equation for the systems with dissipative processes, Journal of Statistical Mechanics: Theory and Experiment, (2017) № 053207
33. Loup Verlet, Computer «Experiments» on Classical Fluids. I. Thermodynamical Properties of Lennard-Jones Molecules, Phys. Rev. 159, 98 – Published 5 July 1967
34. Smirnov N., «Table for estimating the goodness of fit of empirical distributions», Annals of Mathematical Statistics (1948) **19**: 279–281.
35. Kolmogorov A., «Sulla determinazione empirica di una legge di distribuzione», G. Ist. Ital. Attuari. (1933) **4**: 83–91
36. Simard R., L'Ecuyer P., «Computing the Two-Sided Kolmogorov-Smirnov Distribution», Journal of Statistical Software (2011) **39** (11): 1–18
37. David, H. A.; Gunnink, Jason L., «The Paired t Test Under Artificial Pairing», The American Statistician (1997) 51 (1): 9–12.

38. Sawilowsky, Shlomo S., «Misconceptions Leading to Choosing the t Test Over The Wilcoxon Mann–Whitney Test for Shift in Location Parameter», Journal of Modern Applied Statistical Methods (2005) 4 (2): 598–600.
39. Wasserstein, Ronald L.; Lazar, Nicole A., The ASA's Statement on p-Values: Context, Process, and Purpose, The American Statistician (2016) **70** (2): 129–133.
40. Bhattacharya, Bhaskar; Habtzghi, Median of the p value under the alternative hypothesis, The American Statistician (2002). **56** (3): 202–6.
41. Fisz, Marek, Significance Testing. Probability theory and mathematical statistics (3 ed.). New York: John Wiley and Sons, Inc. (1963) p. 425.
42. R. Courant and D. Hilbert, Methods of mathematical physics. Partial differential equation. vol.2, 1962, New York, London.